*Original Article*

# Enterprise API Transformation: Driving towards API Economy

Naga Mallika Gunturu

[1]*IEEE member, IEEE Computer Society Member, Virginia USA*

***Abstract -*** *API proliferation is expected to grow in the coming years. This growth is further increased by the recent trends in digital transformation efforts undertaken by organizations across the spectrum. This paper discusses the benefits organizations can leverage by moving towards an API economy and proposes a framework for organizations to execute API transformation successfully.*

## 1. Introduction
Organizations across domains strive to improve their business agility, and the COVID pandemic has only increased the urgency. In simple terms, business agility indicates the ability of the organization to adapt and adjust to the changing business environment, so they have a higher potential to seize opportunities and consequently thrive and succeed. To survive and succeed in the current VUCA (Volatile, Uncertain, Complex, and Ambiguous) times, it is natural that organizations across the spectrum are striving to deploy agility at scale. A key aspect to note is that technical agility is essential to achieving the desired levels of business agility [1]. Technical agility is adapting quickly and smoothly or integrating current technologies with new, disruptive, expansive, or convergent technologies [2]. In this context, the Application Program Interfaces (APIs) are gaining momentum rapidly. As a technical concept, an API is a set of rules and protocols for building and integrating application software. APIs let your product or service communicate with other products and services without having to know how they're implemented[3]

Though APIs have been around for a while, the recent explosion in the enterprise digital transformation needs has propelled API usage to the forefront. According to a recently published API Management Market Research Report, the global API market size will grow from US $4.1 billion in 2021 to the US $8.41 billion in 2027, recording a CAGR of about 34% during the forecast period [4].

## 2. Role of APIs in corporate digital transformation
### *2.1. Connected customer experience*
Mulesoft reports that though enterprises are doubling down on the digital transformation efforts, 54% of consumers reported not experiencing a seamless customer journey and generally felt that retail teams across sales, service and marketing don't share information. These data silos result from disconnected systems, many of which rely on aging technology and legacy infrastructure, making the idea of building a connected retail experience seem impossible [15]. A connected experience requires an integration strategy across the entire retail value chain. APIs enable such an integration paving a solid road for seamless customer digital journeys.

### *2.2. Foundation for hyper-automation*
The Mulesoft report referred to above also states that traditional integration takes time. In the current fast-paced world, organizations are heavily challenged to optimally utilize the limited resources - human and infrastructure- to derive maximum benefits. However, benefits maximization is not always possible as often resources are spent on mundane but important manual activities. In this context, APIs help automate processes and free up valuable resources to focus on more value-oriented activities. Expanding this concept of automation to an enterprise-level brings us hyper-automation. There are also significant cost savings enabled by hyper-automation. Gartner forecast that, by 2024, hyper-automation will allow organizations to lower operational costs by 30%. This is crucial in today's digital world, enabling organizations to keep pace with competitors and offer valuable experiences and services to their customers [16].

### *2.3. Increased agility*
Agility gains provided by APIs are two-fold. Laying a foundation to automation enables the resources to be more fungible and stay focused on high-value initiatives. This leads to increased agility and enables organizations to reduce time to market and delight their customers with more frequent releases. On a more low level, as APIs abstract out

Naga Mallika Gunturu / IJCTT, 70(6), 44-50, 2022

how specific functionality is implemented, it enables IT teams to embrace Continuous Integration and Continuous Delivery principles and easily embark and mature their agile delivery processes.

## 3. Challenges with the proliferation of APIs

A recent report published by Gartner [25] emphasizes that "Like the two sides of a coin, APIs provide software engineering leaders with key enabling capabilities to improve their technical posture and business models; but on the flip side, APIs create management and security challenges (see Figure 1)."

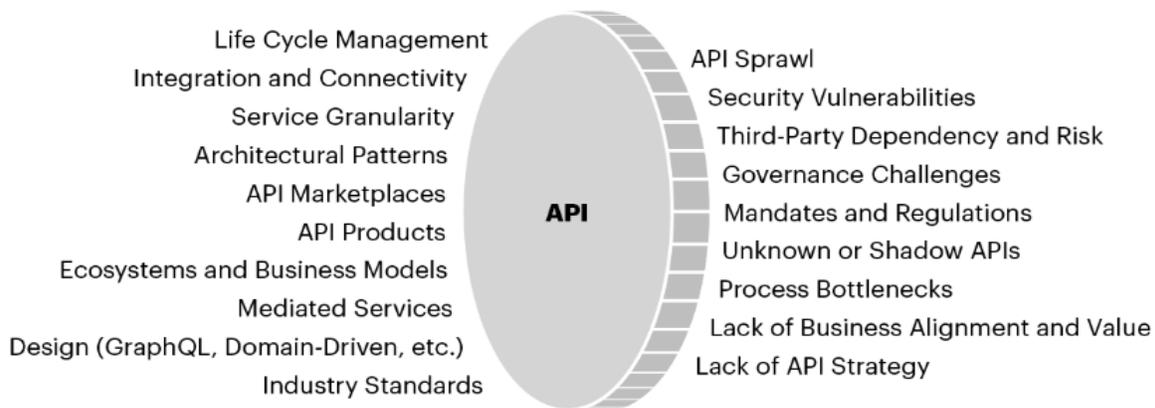

Figure 1: The Two Sides of the API Coin

Before delving into how to mitigate the above challenges effectively and have a successful enterprise-level API transformation, let us look into the key success factors.

### 3.1. Executive agreement and sponsorship

An Enterprise-wide API transformation program's key benefit is to drive business agility. Initiatives of this nature are transversal and need engagement and contributions from several departments - IT, Product Management, Finance, Legal, Security, and Marketing, to name a few. Recognising that a successful API transformation is not just technically plumbing services together but needs the right mindset. By this, we mean having a product-driven mindset that focuses on customer value creation, time to market, and constant improvement. Organizations must be willing to break through barriers from the teams or management and move forward. Hence, strong executive buy-in and sponsorship are essential to have support from all fronts.

### 3.2. API strategy

As mentioned earlier, APIs have been in existence for a while as a technology to glue software programs together. However, as enterprises become technology-driven organizations fuelled by the benefits of APIs, it is important to have a clear API strategy that helps them identify and act on the best API opportunities. It addresses the vital question of why the organization needs the API transformation initiative and provides the vision, mission, high-level approach, and the limitations to which the transformation should adhere. This strategy acts as a guiding north star to the teams throughout the transformation.

### 3.3. Security and compliance

Businesses of any nature must adhere to regulatory compliance requirements. Failure to meet regulatory requirements will negatively impact customers' trust in a business [1]. API transformation brings in a set of new risks as data is easily exposed to outside entities, thus necessitating strict data classification and access controls.

Also, several legal aspects must be considered depending on the API business model, especially in established partnerships. Addressing these risks through strong governance principles is essential to avoid financial and reputational losses.

### 3.4. Change management
Like all major transversal initiatives, API transformation brings many changes: technology, business, people, and processes. An effective change management process should be implemented to ensure people, processes, and business transformation to the new way of working with minimum impact. [1].

## 4. API Transformation Framework
Enterprise API transformation is an effort-intensive strategic initiative that can typically stretch into multi-year roll-outs. While extensive information is available on factors organizations should consider to decide upon the API business model and the associated governance considerations, there is not a lot of material available on the practical aspects that would enable organizations to launch and successfully execute API transformation programs. This paper proposes the following 4-phased framework to address this gap.

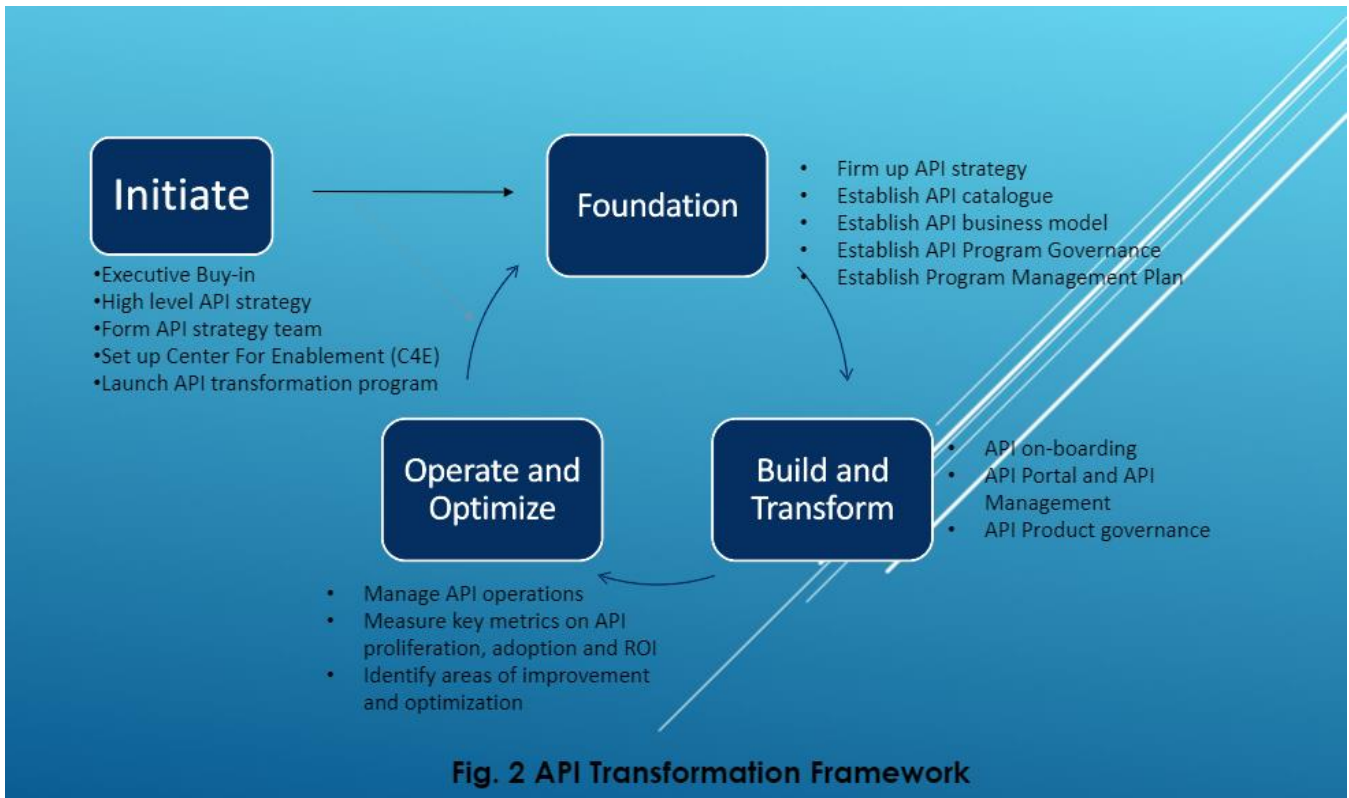

Fig. 2 API Transformation Framework

### 4.1. Phase 1 – Initiate
The first step for an enterprise API transformation is to have a solid assessment to formulate its API strategy. This phase's key outcome is alignment at the executive level on the API transformation's why what, and how. Key elements of the API transformation initiative covering the high-level approach and limitations within which the transformation should happen are established. Key questions that help drive this assessment are:
a) How does API transformation contribute to the organization's strategy?
b) What does the organization stand to lose by not investing in API transformation?
c) What purpose and key benefits is the organization expecting to reap from API transformation?
d) How ready is the organization for API transformation regarding people, processes, and tools?
e) What are the risks involved?
f) What complexity is involved in executing API transformation given the current state vis-a-vis the target future state?
g) What business or technical processes should be adapted as a priority?

Addressing the above questions and establishing a common consensus among executive leadership is essential to building a strong business case that helps establish buy-in for API transformation from teams across the organization. It is important to note that API transformation requires organizations to not only adapt on the technical front but should also be prepared to adapt the organization's mindset and culture, and processes to embrace a mindset of value-driven delivery revolving around customer centricity and continuous improvement.

After establishing executive agreement and sponsorship, the next step is to form the teams which would be responsible for the successful execution of the transformation objectives:

1. To have sustained executive engagement, an API strategy team whose focus is to maintain alignment between business priorities and API transformation efforts should be formed. This team should have participants in product management, IT, Operations, Legal, Security, and architecture.
2. Establish a Center for Enablement (C4E). One of the key aspects of an API transformation program is that it brings in a fundamental shift in the operating model. C4E is a cross-functional team of resources from IT lines of business whose mandate is to productize, publish, and harvest reusable assets and best practices. They promote consumption and collaboration and help drive self-reliance while improving results through feedback and metrics [5].
3. An enterprise-wide program with the following workstreams to manage API transformation is launched following workstreams.

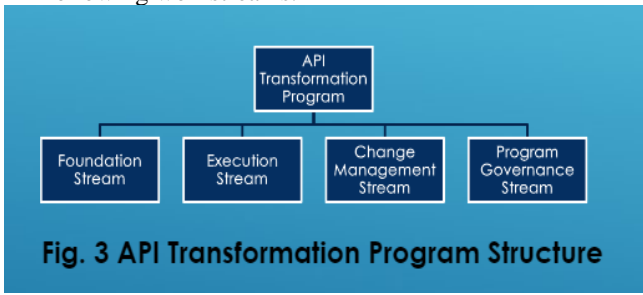

Fig. 3 API Transformation Program Structure

*4.1.1. Foundation stream*
Set up the basic building blocks so the API transformation can start and proceed smoothly. This phase also focuses on defining the necessary guidelines and guardrails that determine how the transformation will be rolled out.

*4.1.2. Execution stream*
Focus is the creation and management of APIs

*4.1.3. Change Management stream*
Prepare and adapt people and processes to the mindset and organizational culture that is conducive to the API transformation

*4.1.4. Program Governance stream*
Focus is on delivering the program objectives within the established constraints. This workstream spans all phases of the program, starting with establishing program organization and later during the execution phase focussing on assessing, managing, and reporting the progress.

The program kick-off meeting marks the launch of the program. The program manager conducts this meeting with the participation of the key stakeholders - the API strategy team, C4E, development and operations leads, and support teams like vendor management, legal, and procurement.

*4.2. Phase 2 – Foundation*
As it is self-explanatory, the focus of this phase is to establish the foundation blocks needed to support the transformation by carrying out various activities of the Program Governance and Foundation workstreams.

*4.2.1. Program Governance workstream*
At this stage, this workstream aims to put in place several building blocks needed to "manage the program" - communication plan, change management plan, schedule management plan, RACI, reporting framework, risk and issue log.

Program Governance workstream will start small and expand as outputs of the other workstreams are delivered. For instance, at the start, vendor management is not part of the program governance, but once a decision is taken on the API business model and the associated relationships with vendors and partners are established, a plan for managing the related vendor procurement and management activities will be added to the program management plan.

*4.2.2. Foundation workstream*
*Assess current state*
The C4E team and the different delivery teams assess the infrastructure and the APIs already available in the organization. Using this information, they should establish a catalogue of APIs. The next step is to identify which of the existing APIs are important and where there is a need to expose certain data elements or services as an API. At this stage, some high-level thoughts on the API granularity are also established, along with any trade-offs needed to balance achieving program goals and programmatic ambitions.

*Define API strategy*
In parallel to evaluating the current state, the C4E team, in collaboration with the API strategy team, should also work to define the API strategy covering the API business model and API governance principles that the organization should abide by. While the API strategy might get updated and fine-tuned with time, the strategy blueprint will be a guiding north star. Typical questions that would be answered as part of this phase are as follows:

a. What are the intended business outcomes? The direction from the executive leadership as part of the "Initiate" phase is further elaborated here
b. Who is the target audience? What does the target ecosystem look like? What is the role that organization would play in this ecosystem?
c. What is the value proposition between the organization and the other ecosystem participants? This leads to identifying the appropriate API business model(s) that the organization would employ.
d. What assets need to be exposed via APIs, and what is the priority for these?

### 4.3. API ecosystem
Typically there are 4 categories of stakeholders in an API ecosystem:

#### 4.3.1. API Provider
The organization that exposes data and capabilities through an Application Programming Interface (API) or a service. Such a provider can be considered to be an API Platform when it enables consumers to create shared value[6]

#### 4.3.2. API consumer
The stakeholder who integrates APIs and builds software applications or websites.

4.3.3. API customer
The stakeholder who takes the ultimate decision to commercially pay for the API services provided by the provider and consumer

#### 4.3.4.API end-user
The stakeholder representing those who use the applications or the websites that use APIs.

### 4.4. API Business model
Organizations establish business models to define two important aspects - "Part one includes all the activities associated with making something: designing it, purchasing raw materials, manufacturing, and so on. Part two includes all the activities associated with selling something: finding and reaching customers, transacting a sale, distributing the product, or delivering the service. A new business model may turn to design a new product for an unmet need or a process innovation. That is, it may be new in either end." [9]

The following questions need to be considered to determine the appropriate monetization model:[10]
1. What is the true value of the assets being exposed?
2. Who considers this asset to be valuable? - Is it an API consumer or the API customer?
3. Who owns the relationship with the end-user?
4. Will enforcing a price on the availability of the asset act as a deterrent for consumers and customers to access the APIs?
5. Does tiered pricing make sense?

Four main models for monetization of APIs are widely prevalent.[10].

#### 4.4.1. Free
In this model, the consumer and customer are not charged any fee for consuming APIs. Typical scenarios when this model is used: To drive adoption of APIs, To make headway into new channels, To enhance brand loyalty, basically when exposing the asset is a low cost compared to the benefits it returns.

#### 4.4.2. Developer Pays
In this model, the consumer of the APIs pays the provider for access to the APIs. The consumer could further pass these charges to the API customers and the end-users. Several sub-categories of this model are relevant for a broad category of scenarios - Pay As You Go, Freemium, Tiered, Points-Based, and Transaction Fee.

#### 4.4.3. Developer gets Paid
In this model, the consumer of the APIs gets paid by the API provider for API consumption. That is, API producers, incentivise the consumers to their API services. Several sub-categories of this model exist relevant to a broad category of scenarios - Revenue Share, Affiliate, and Referral

#### 4.4.4. Indirect
In this model, APIs are not a direct revenue source but fulfill purposes leading to revenue generation. The different sub-categories of this model are Content Acquisition, Content Syndication, Internal—Consumer, Internal—Non-consumer, B2B Customer, B2B Partner, and Business Expansion.

An organization often adopts API transformation for non-monetization purposes to achieve objectives like legacy modernization. In this case, the organization can establish a chargeback policy with which the consumer division "pays" the producer division.

It is also worth noting that sometimes organizations partner with other API or data providers to create new assets that are valuable to API consumers and customers.

### 4.5. API Governance
API governance is a key differentiator for any organization's competitive edge. The technical and business synergies realized through proactive management of a company's API architecture are too great to be ignored [11]. A proper API governance model ensures that the APIs are discoverable, follow consistent principles, follow the organization's security policies, and are truly reusable. API governance should balance business drivers and requirements, risks, and the existing compliance requirements. The existing policies and security practices must be extended for the API paradigm [1].

### 4.6. Phase 3 - Build / Transform

Focus during this phase is to build the APIs according to the roadmap and evolve the API management practices in the organization through the program change management, foundation, and execution streams.

### 4.7. Change Management workstream

The focus of C4E in this phase is to evangelise APIs within the organization and drive their consumption within the organization. Key factors that help the spread of API usage and mindset in the organization are:

1. Provide simple and reliable tools and processes with which developers can develop and integrate with the APIs
2. Conduct brown bag sessions and communicate the value of APIs within the organization
3. Engage with early adopter teams and work closely with them, taking in their feedback and improving the processes, tools, frameworks, and policies and showcasing it to other teams.
4. Encourage and cultivate champions in all teams.

### 4.8. Foundation workstream

A proactive approach to API management is needed to ensure that the organization is well prepared to manage the growth in the usage and the number of APIs produced by the organization. In this phase, the foundation work stream focuses on setting up API Portal and API Gateway, which provide API management. The API Portal is for developers to sign up to use APIs and receive API Keys and quotas. The API Gateway operates at runtime, managing the API Key usage and enforcing the API usage quotas. The API Gateway also performs the very important task of bridging from the technologies used by API clients (REST, OAuth) to the technologies used in the enterprise (Kerberos, SAML, or proprietary identity tokens such as CA SiteMinder sm session tokens[12]. The key aspects to remember are simplicity and ease of use. Ensuring that developers realise the ease with which APIs can be consumed is an essential success factor for API adoption.

### 4.9. Execution workstream

This phase focuses on building APIs and exposing the assets in the order of priority defined by the API strategy. This phase also focuses on enhancing the existing APIs to the latest governance principles. In this phase, a product-specific governance model is also established. API product governance centers around the API lifecycle management and helps address the following key questions[13]:

1. Who owns the product life cycle overall?
2. How is the product lifecycle managed?
3. How well-aligned are the product vision, business model, market strategy, product design, roadmap, and operating model?
4. How effectively is new feature delivery balanced with the retirement of technical debt?
5. What channels (automated and user-based) are in place to collect consumer feedback and measure API product performance?
6. What API product risks need to be managed, and what regulatory compliance is required?

### 4.10. Phase 4 - Operate and Optimize

The program team receives very important feedback and lessons learned from the experiences of different teams as they onboard the API journey. The focus of this phase should be to optimize the processes and tools so that API development and consumption continue to increase, contributing to the success of the API transformation program. Further, key metrics around API proliferation, utilization, and ROI should be measured during this phase, which provide insights and inputs to the Foundation and Change management streams.

## 5. Conclusion

The proposed 4-phase framework weaves the critical success factors into the transformation program layout. By deploying this framework, organizations can ensure the success of their API transformation initiatives and fast-track their digital transformation by leveraging the full benefits that the API economy offers.

## References
[1] Naga Mallika Gunturu , "A Framework for Successful Corporate Cloud Transformation," *International Journal of Computer Trends and Technology*, vol. 70, no.3, pp. 9-15, 2022. https://doi.org/10.14445/22312803/IJCTT-V70I3P102
[2] The ComputerWorld website, 2020. [Online]. Available: https://www.computerworld.com/article/2582993/technical-agility.html
[3] The Redhat website, 2002. [Online] Available: https://www.redhat.com/en/topics/api/what-are-application-programming-interfaces
[4] The Marketdataforecast website, 2002.[Online] Available: https://www.marketdataforecast.com/market-reports/api-management-market
[5] The Mulesoft website, 2018.[Online] Available: https://blogs.mulesoft.com/digital-transformation/it-management/what-is-a-center-for-enablement-c4e/

[6] The Nordicapis website, 2014.[Online] Available: https://nordicapis.com/api-platform-defined-api-provider-is-a-platform/
[7] The IBM website, 2016.[Online] Available: https://www.ibm.com/downloads/cas/L5Q82XR0
[8] The dicentral website, 2021. [Online] Available: https://api.dicentral.com/api-stretegy
[9] The HBR website, 2015.[Online] Available: https://hbr.org/2015/01/what-is-a-business-model
[10] The IBM website, 2016.[Online] Available: https://www.ibm.com/downloads/cas/L5Q82XR0
[11] The Forbes website, 2020.[Online] Available: https://www.forbes.com/sites/marksettle/2020/05/04/cios-weaponize-apis-through-better-governance/?sh=23057d206076
[12] The Dzone website, 2013.[Online] Available :https://dzone.com/articles/api-gateway-and-api-portal
[13] The Mulesoft website, 2020. [Online] Available: https://blogs.mulesoft.com/api-integration/strategy/4-ps-of-api-governance
[14] Akshay Heroor "Core System Modernization" *International Journal of Computer Trends and Technology*, vol. 68, no.11 pp.73-78, 2020.
[15] The Mulesoft website, 2021. [Online] Available: https://www.mulesoft.com/sites/default/files/resource-assets/eBook%20-%20Building%20the%20connected%20retail%20experience.pdf
[16] The Techradar website, 2021. [Online] Available: https://www.techradar.com/features/using-api-integration-to-turn-automation-into-hyperautomation
[17] The Intellias website, 2020. [Online]. Available: https://intellias.com/the-api-strategy/
[18] The IBM website, 2020. [Online] Available: https://www.ibm.com/cloud/learn/api
[19] The Postman website, 2021. [Online] Available: https://www.postman.com/state-of-api/
[20] The Softwareag website, 2021. [Online] Available : https://www.softwareag.com/en_corporate/resources/asset/ar/integration-api/apis-integration-microservices-report.html
[21] The Apigee website, 2021. [Online] Available: https://pages.apigee.com/rs/351-WXY-166/images/Apigee_StateOfAPIS_eBook_2020.pdf
[22] The Deloitte website, 2015. [Online] Available : https://www2.deloitte.com/content/dam/Deloitte/us/Documents/financial-services/us-fsi-api-economy.pdf
[23] The Techtarget website, 2021. [Online] Available: https://www.techtarget.com/searchapparchtecture/Guide-to-building-an-enterprise-API-strategy
[24] The ITProportal website, 2019. [Online] Available: https://www.itproportal.com/features/the-api-ecosystem-what-it-leaders-need-to-know/
[25] The Gartner website, 2021.[Online] Available: https://www.gartner.com/doc/reprints?id=1-28X5Q6BX&ct=220127&st=sb&aliId=eyJpIjoidEE4cEhJdUk0YWtGTEdhVyIsInQiOiJVMENCUXgybko5aUd2K0NtOFFaNUJRPT0ifQ%253D%253D